\documentclass[showpacs,twocolumn,pra]{revtex4}

\usepackage{epsf}
\usepackage{amsmath}
\usepackage{amsfonts}
\usepackage{amssymb}
\usepackage{bm}
\usepackage{bbm}
\usepackage{nicefrac}
\usepackage{color}
\usepackage{maybemath}
\usepackage{pifont}

\usepackage{graphicx}
\usepackage{epsfig}
\usepackage{latexsym}
\usepackage{dcolumn}
\usepackage{pifont}

\def\ii{{\mathrm{i}}}

\newcolumntype{.}{D{x}{}{-1}}

\begin{document}

\newcommand{\addrROLLA}{Department of Physics,
Missouri University of Science and Technology,
Rolla, Missouri 65409-0640, USA}

\title{Self--Energy Correction to the Bound--Electron 
$\maybebm{g}$ Factor of $\maybebm{P}$ States}

\author{Ulrich D. Jentschura}
\affiliation{\addrROLLA}

\bibliographystyle{myprsty}

\begin{abstract}
The radiative self-energy correction to the bound-electron $g$ factor of 
$2P_{1/2}$ and $2P_{3/2}$
states in one-electron ions is evaluated to order $\alpha \, (Z\alpha)^2$.
The contribution of high-energy virtual photons is treated by means of an
effective Dirac equation, and the result is verified by an 
approach based on long-wavelength quantum electrodynamics. 
The contribution of low-energy virtual photons is calculated both 
in the velocity and in the length gauge and gauge invariance is verified explicitly.
The results compare favorably to recently available numerical data
for hydrogenlike systems with low nuclear charge numbers.
\end{abstract}


\maketitle

%
%
\section{Introduction}

When a bound electron interacts with an external, uniform and 
time-independent magnetic field (Zeeman effect), the energetic degeneracy of the 
atomic energy levels with respect to the 
magnetic projection quantum number is broken, and the 
different magnetic sublevels split according to the formula
\begin{equation}
\Delta E = g_j \, \mu_B \, B \, \mu
\end{equation}
where $g_j$ is the bound-electron (Land\'{e}) $g$ factor,
$\mu_B = -e/(2m)$ is the Bohr magneton, and $B$ is the magnetic field
which is assumed to be oriented parallel to the quantization axis. Finally, 
$\mu$ is the magnetic projection quantum number of the 
electron; i.e.~the projection of its total angular momentum 
(divided by $\hbar$) onto the quantization axis.

In leading order, the bound-electron $g$ factor is determined
by nonrelativistic quantum theory and is equal to a 
rational number for all bound states in a hydrogenlike ion.
Both relativistic atomic theory as well as
quantum electrodynamics (QED) predict deviations from the 
nonrelativistic result. The relativistic effects follow
from Dirac theory and can be expressed in terms 
of a power series in the parameter $Z\alpha$, where 
$Z$ is the nuclear charge number and $\alpha$ is the fine-structure 
constant. The QED effects are caused mainly by the anomalous magnetic 
moment of the electron, which is turn in caused by the exchange
of high-energy virtual photons before and after the interaction 
with the external magnetic field.
Here, by ``high-energy'' we refer to a virtual photon with an energy of 
the order of the electron rest mass.
A second source for QED effects are exchanges of virtual photons
with an energy commensurate with the atomic binding energy scale,
which is smaller than the electron rest mass energy 
by a factor $(Z\alpha)^2$. Here, the electron emits and 
absorbs a virtual photon before and after the interaction with the 
external magnetic field, undergoing a 
virtual transition to a excited atomic state in the middle. 
For $P$ states, the latter effects lead to a correction to the 
bound-electron $g$ factor of order $\alpha (Z\alpha)^2$. 
The complete result for the correction of order $\alpha (Z\alpha)^2$
is obtained after adding the anomalous magnetic moment correction 
(high-energy part) and the low-energy photon contribution of the 
same order.

Previous studies of the bound-electron $g$ factor for $P$ states in
hydrogenlike systems include
Refs.~\cite{BrPa1967,He1973,GrKa1973,GrHe1973,CaGrOw1978}. 
Quite recently, the problem has received renewed
interest~\cite{AnSe1993,GoSa1997}. For few-electron
ions, the bound-electron $g$ factor has been investigated in
Refs.~\cite{AnSe1993,GoSa1997,YaDr1994,Ya2002,Pa2004}.
For the $2^3 P$ states of helium,
there is still an unresolved discrepancy of theoretical 
and experimental results
(see Refs.~\cite{LeHu1973,LhFaBi1976,Pa2004}).

The expansion of the quantum electrodynamic radiative 
correction to the electron $g$ factor, which is an
expansion in powers of $\alpha$ for a free electron,
is intertwined with an expansion in powers of $Z\alpha$ for a 
bound electron (this fact has been stressed in Ref.~\cite{JeEv2005}). 
For an $nP_j$ state in a hydrogenlike system,
we can write down the following 
intertwined expansion in powers of $\alpha$ and $Z\alpha$, 
\begin{align}
\label{defcoeff}
\delta g(nP_j) =& \; g_{00} + 
(Z\alpha)^2 \, \frac{g_{20}}{n^2} +
{\mathcal O}(Z\alpha)^4
\nonumber\\[2ex]
& \; + \frac{\alpha}{\pi} \left\{ b_{00} + 
(Z\alpha)^2 \, \frac{b_{20}}{n^2} + 
{\mathcal O}(Z\alpha)^4 \right\} \,.
\end{align}
The coefficients $g_{00}$ and $g_{20}$ characterize the 
relativistic effects, whereas $b_{00}$ and $b_{20}$ are obtained 
from the one-loop radiative correction.
The nonrelativistic result for the Land\'{e} $g$ factor reads
\begin{equation}
\label{g00}
g_{00}(nP_{1/2}) = \frac23 \,, \qquad
g_{00}(nP_{3/2}) = \frac43 \,.
\end{equation}
The relativistic correction follows from Breit 
theory and the Dirac equation in an external magnetic 
field~\cite{Br1928,GrKa1973},
\begin{equation}
\label{g20}
g_{20}(nP_{1/2}) = -\frac{2}{3} \,, \qquad
g_{20}(nP_{3/2}) = -\frac{8}{15} \,.
\end{equation}
The leading correction due to the anomalous magnetic 
reads as (see Refs.~\cite{BrPa1967,GrKa1973}),
\begin{equation}
\label{b00}
b_{00}(nP_{1/2}) = -\frac13 \,, \qquad
b_{00}(nP_{3/2}) = \frac13 \,.
\end{equation}
We are concerned here with the evaluation of the 
$b_{20}$ coefficient of $nP_j$ states,
which is determined exclusively by self-energy 
type corrections (vacuum polarization does not contribute).

We adopt the following outline for this paper.
In Sec.~\ref{hep}, we reexamine the contribution 
of high-energy virtual photons (see also Ref.~\cite{GrKa1973}).
Two alternative derivations are presented, which are based on an effective
Dirac equation (Sec.~\ref{hep1}) and on an 
effective low-energy long-wavelength quantum electrodynamic 
theory (Sec.~\ref{hep2}) which is obtained 
from the fully relativistic theory by 
a combined Foldy--Wouthuysen and Power--Zienau transformation~\cite{Pa2004}.
The low-energy part is also treated in two alternative ways.
The velocity-gauge calculation in Sec.~\ref{lep1} is contrasted
with the length-gauge derivation in Sec.~\ref{lep2}.
Conclusions are reserved for Sec.~\ref{conclu}.
Natural units ($\hbar = c = \epsilon_0 = 1$) are
used throughout the paper.

%
%
\section{High--Energy Part}
\label{hep}

%
%
\subsection{Effective Dirac Equation}
\label{hep1}

In Ref.~\cite{GrKa1973}, the contribution to 
$b_{20}$ due to high-energy virtual photons was obtained 
on the basis of the two-body Breit Hamiltonian.
Here, we perform the calculation using a simple approach,
based on an effective Dirac Hamiltonian (see Ch.~7 of Ref.~\cite{ItZu1980}).
For an electron interacting with external electric and magnetic
fields, this equation reads
\begin{align}
\label{HDm}
H_{\rm rad} =& \; \vec{\alpha} \cdot
\left[\vec{p} -{\mathrm e} \, F_1(\vec{\nabla}^2) \, \vec{A}\right]
+ \beta\,m + F_1(\vec{\nabla}^2) \, V
\nonumber\\
& + F_2(\vec{\nabla}^2) \, \frac{e}{2\,m} \, \left({\mathrm i}\,
\vec{\gamma} \cdot \vec{E} - \beta \, \vec{\Sigma} \cdot \vec{B}
\right)\,.
\end{align}
We here take into account the
Dirac form factor $F_1$ and the Pauli form factor $F_2$.
The matrices $\vec{\alpha} = \gamma^0 \vec{\gamma}$ 
and $\beta = \gamma^0$ are the standard Dirac matrices
in the Dirac representation~\cite{ItZu1980},
$m$ is the electron mass,
and $e = - |e|$ is the electron charge.
Up to the order relevant for the current calculation,
we may approximate both form factors in the limit of 
vanishing momentum transfer as
\begin{equation}
F_1(\vec{\nabla}^2) \approx F_1(0) = 1 \,, \quad
F_2(\vec{\nabla}^2) \approx F_2(0) \approx \kappa \equiv \frac{\alpha}{2\pi} \,.
\end{equation}
The vector potential $\vec{A}$ corresponds to a uniform 
external magnetic field, i.e.~$\vec{A} = 
\tfrac12 \, \left( \vec{B} \times \vec{r} \right)$, and 
the electric field $\vec{E}$ is that of the Coulomb potential
($e\, \vec{E} = - \vec{\nabla} V$). Finally,
$V = -Z\alpha/r$ is the binding potential. So,
\begin{align}
\label{Hrad}
H_{\rm rad} \approx& \; \vec{\alpha} \cdot \vec{p} 
+ \beta\,m + V
-\frac{e}{2} \, \vec{\alpha} \cdot \left( \vec{B} \times \vec{r} \right)
\nonumber\\[2ex]
& \; 
-\frac{\ii \kappa}{2 m} \, \vec{\gamma} \cdot \vec{\nabla} V
-\frac{e}{2 m} \kappa \, \beta \, \vec{\Sigma} \cdot \vec{B}  \,.
\end{align}
Dirac eigenstates fulfill $(\vec{\alpha} \cdot \vec{p} + \beta\,m + V) \psi =
E_D \psi$, where $E_D$ is the Dirac energy. The first few
terms in the perturbative expansion of $E_D$ in a magnetic field read
\begin{align}
\label{dirac}
\Delta E =& \;
\left< \psi \left| 
-\frac{e}{2} \, \vec{\alpha} \cdot \left( \vec{B} \times \vec{r} \right)
\right| \psi \right>
\\[2ex]
& \; -\frac{e}{2 m} \kappa \, 
\left< \psi \left| 
\beta \, \vec{\Sigma} \cdot \vec{B} 
\right| \psi \right> 
\nonumber\\[2ex]
& \; + \frac{e \, \kappa}{2 m}  \, 
\left< \psi \left| 
\left( \ii \vec{\gamma} \cdot \vec{\nabla} V \right) 
\frac{\cal Q}{2 m} 
\left[ \vec{\alpha} \cdot (\vec{B} \times \vec{r} ) \right]
\right| \psi \right> \,.
\nonumber
\end{align}
where ${\cal Q} = \tfrac12 \, (1 - \gamma^0)$ is a projector onto 
virtual negative-energy states, and $\psi$ is the 
relativistic wave function. 
An evaluation of the first term on the right-hand side 
of~\eqref{dirac} with Dirac wave functions
confirms the results for $g_{00}$ and $g_{20}$ given in
Eqs.~\eqref{g00} and~\eqref{g20}.
The second term on the right-hand side of~\eqref{dirac} 
yields the result for $b_{00}$ as given in Eq.~\eqref{b00}.
When the Dirac wave functions are properly expanded in
powers of $Z\alpha$, the second and third terms
on the right-hand side of~\eqref{dirac} yield the following
high-energy contribution $b^{(H)}_{20}$
to the $b_{20}$ coefficient defined in Eq.~\eqref{defcoeff},
\begin{equation}
\label{bH20}
b^{(H)}_{20}(nP_{1/2}) = - \frac{1}{2} \,, \qquad
b^{(H)}_{20}(nP_{3/2}) = \frac{1}{10} \,.
\end{equation}
These results are in agreement with those given in Eq.~(5) 
of Ref.~\cite{GrKa1973}.

%
%
\subsection{Long--Wavelength Quantum Electrodynamics}
\label{hep2}

It is instructive to compare the fully relativistic 
approach outlined above to an effective nonrelativistic theory.
In Ref.~\cite{Pa2005}, a systematic procedure has been described in order
to perform a nonrelativistic expansion of the interaction Hamiltonian 
for a light atomic system with slowly varying external electric and magnetic
fields. This procedure involves two steps, (i) a Foldy--Wouthuysen transformation
of an interaction of the type~\eqref{HDm}, suitably generalized for many-electron
systems, and (ii) a Power--Zienau transformation to express the 
vector potentials in terms of physically observable field strengths.
The result is an interaction, given in Eq.~(30) of Ref.~\cite{Pa2005}, 
which describes a nonrelativistic expansion of the atom-field 
interaction in powers of $Z\alpha$ and can be used in order to 
identify terms which contribute at a specified order.

If we are interested in evaluating the corrections
to the $g$ factor up to order $\alpha (Z\alpha)^2$, 
i.e.~all corrections listed in Eq.~\eqref{defcoeff},
the relevant effective interactions for a one-electron system are
\begin{subequations}
\label{Hmag}
\begin{align}
\label{H1mag}
H_{\rm mag} =& \; H_M + \sum_{i=1}^3 H_i  \,, \quad
H_M = \mu_B ( \vec{L} + \vec{\sigma} ) \cdot \vec{B} \,,
\\[2ex]
H_1 =& \; - \frac{\mu_B}{2 m^2} \,
\vec{p}^{\,2} \, ( \vec{L} + \vec{\sigma} ) \cdot \vec{B} \,,
\\[2ex]
H_2 =& \; \frac{\mu_B (1 + 2\kappa)}{4 m} 
\frac{Z\alpha}{r^3} \,
(\vec{r} \times \vec{\sigma})  \cdot
(\vec{r} \times \vec{B})  \,,
\\[2ex]
H_3 =& \; -\frac{\mu_B \kappa}{2 m^2} 
(\vec{p} \cdot \vec{\sigma}) 
(\vec{p} \cdot \vec{B}) \,,
\end{align}
\end{subequations}
where $\mu_B = -e/(2 m)$ is the Bohr magneton.
We denote the Schr\"{o}dinger--Pauli two-component wave
function by $\phi$ in order to distinguish it from the 
fully relativistic wave function $\psi$. 
Specifically, $\phi$ reads as $\phi(\vec r) = R(r) \, 
\chi^\mu_\kappa(\hat r)$ in the coordinate representation,
where $R(r)$ is the nonrelativistic radial component of the wave function and 
$\chi^\mu_\kappa(\hat r)$ is the 
standard two-component spin-angular function~\cite{VaMoKh1988}.
An evaluation of the perturbation
\begin{equation}
\Delta E = \langle \phi | H_{\rm mag}  | \phi \rangle
\end{equation}
confirms the results of Eqs.~\eqref{g00},~\eqref{g20},~\eqref{b00} and~\eqref{bH20}
for the high-energy part.
No second-order effects need to be considered in this formalism up to 
the order in the $Z\alpha$-expansion relevant for the current study.

%
%
\section{Low--Energy Part}
\label{lep}

%
%
\subsection{Velocity Gauge}
\label{lep1}

The most economical approach to the calculation of the 
low-energy contribution of order $\alpha (Z\alpha)^2$
to the $g$ factor of $P$ states 
consists in a calculation of the orbital
$g_\ell$ factor, with a conversion of the orbital 
$g_\ell$ factor to the Land\'{e} $g_j$ factor in a second 
step of the calculation. In the order $\alpha(Z\alpha)^2$, 
one may indeed convert the
spin-independent correction to the orbital $g_\ell$ factor 
to a spin-dependent correction to the 
$g_j$ factors of the $2P_{1/2}$ and $2P_{3/2}$, as
described in Appendix~\ref{appa}.
However, a more systematic 
approach to the problem, which is also applicable 
to higher-order (in $Z\alpha$) corrections, 
is based on a perturbation of the nonrelativistic 
self-energy of the bound electron by the 
magnetic interaction Hamiltonian~\eqref{Hmag}.
This is the approach outlined below.

We thus investigate the perturbation of the nonrelativistic
bound-electron self-energy~\cite{Be1947} due to the 
magnetic interaction Hamiltonian
\begin{equation}
H_M = -\frac{e}{2 m} \, 
\left( \vec{L} + \vec{\sigma} \right) \cdot \vec{B} 
\end{equation}
given in Eq.~\eqref{H1mag}.
In the velocity gauge, the interaction of the electron 
with the vector potential $\vec{A}$ of the quantized electromagnetic 
field is given by the term 
$-\vec{p} \cdot \vec{A}/m$, where $\vec{p}$ is the electron momentum. 
In an external magnetic field, it is the physical momentum 
$\vec{p} - \frac{e}{2} \, ( \vec{B} \times \vec{r} )$,
not the canonical momentum $\vec{p}$, which couples to the 
quantized electromagnetic field. This amount to a correction 
$\delta \vec{J}$ to the electron's transition current given by
\begin{equation}
\delta J^i 
= -\frac{e}{2 m} \, \left( \vec{B} \times \vec{r} \right)^i 
= -\frac{e}{2 m} \, \epsilon^{ijk} B^j r^k \,,
\end{equation}
Because of the symmetry of the problem (Wigner--Eckhart theorem),
we may fix the axis of the $B$ field to be along the 
quantization axis ($z$ axis) and the projection of the 
reference state to be $\mu = \tfrac12$. 
This procedure allows one to simplify the angular algebra.
It is inspired by the separation of nuclear and electronic
tensors that are responsible for the hyperfine interaction.
Such a separation has been used in Eq.~(1) of Ref.~\cite{BiPaFFJo1995}
and in Eqs.~(10) and~(11) of Ref.~\cite{YeArShPl2005}.
In the case of the $g$ factor, the magnetic field of the nucleus is replaced by the 
external, homogeneous magnetic field of the Zeeman effect.

We divide out a prefactor $-e/(2 m)$ from both the 
magnetic Hamiltonian $H_M$ and from the current $\delta J^i$ and 
obtain the perturbative Hamiltonian $h_{M,0}$ and the 
scaled current $\delta j^i_0$. In the spherical basis,
this procedure leads to the operators
\begin{equation}
h_{M,0} = L_0 + \sigma_0 \,, \qquad
\delta j^i_0 = \epsilon^{i3k} \, r^k \,,
\end{equation}
which are aligned along the quantization axis 
of the external magnetic field.
The index zero of the operators,
in the spherical basis, denotes the $z$ component
in the Cartesian basis (see Ref.~\cite{VaMoKh1988}).
The following shorthand notation for the atomic states with 
magnetic projection $\mu = \tfrac12$ proves useful,
\begin{equation}
| j \tfrac12 \rangle \equiv | n P_j (\mu = \tfrac12) \rangle \,,
\qquad
\langle h_{M,0} \rangle \equiv
\langle j \tfrac12 | h_{M,0} | j \tfrac12 \rangle \,.
\end{equation}
Finally, we can proceed to the calculation of the 
perturbed self-energy. The nonrelativistic (Schr\"{o}dinger)
Hamiltonian of the atom is 
\begin{equation}
H_{\rm NR} = \frac{\vec{p}^2}{2 m} + V \,,
\end{equation}
and the nonrelativistic self-energy reads
\begin{align}
\label{deltaE}
\delta E =& \; 
- \frac{2 \alpha}{3 \pi m^2} \,
\int\limits_0^{m \epsilon} d \omega \omega \,
\left< j \tfrac12 \left| \vec{p} \,
\frac{1}{H_{\rm NR}  - E_{\rm NR}  + \omega} \,
\vec{p} \right| j \tfrac12 \right>  \,.
\end{align}
The wave-function correction to the self-energy reads
\begin{align}
& \delta E_\psi =
- \frac{4 \alpha}{3 \pi m^2} \,
\int_0^{m\epsilon} d \omega \omega \,
\\
& \times
\left< j \tfrac12 \left| \vec{p} \,
\frac{1}{H_{\rm NR}  - E_{\rm NR}  + \omega} \,
\vec{p} \left( \frac{1}{E_{\rm NR} - H_{\rm NR}} \right)' 
h_{M,0} \right| j \tfrac12 \right> \,.
\nonumber
\end{align}
Here, $[1/(E_{\rm NR} - H_{\rm NR})]'$ is the reduced Green
function, with the reference state being excluded from the sum over 
virtual states. The contribution of virtual $nP_j$ states
(with $j$ being equal to that of the reference state and 
$n \geq 2$)
vanishes because of the orthogonality of the 
nonrelativistic radial wave functions.
The interaction $h_{M,0}$ couples $2P_{1/2}$ and $2P_{3/2}$ 
states, but the contribution of virtual states with different
$j$ as compared to the reference state 
vanishes after angular algebra~\cite{VaMoKh1988}
because the self-energy interaction operator
$\{ \vec{p} [1/(H_{\rm NR}  - E_{\rm NR}  + \omega)] \vec{p} \}$
is diagonal in the total angular momentum. 
Virtual states with different orbital angular momentum
than the reference state are not coupled at all to the reference
state by the action of the perturbative Hamiltonian $h_{M,0}$.
Because all contributions vanish 
individually, we can thus conclude that $\delta E_\psi = 0$.

Hence, we have to evaluate first-order corrections to the Hamiltonian
and to the energy corresponding to the
replacements $H_{\rm NR} \to H_{\rm NR} + h_{M,0}$
and $E_{\rm NR} \to E_{\rm NR} + \langle h_{M,0} \rangle$
in Eq.~\eqref{deltaE}. Furthermore, we have a correction to the current corresponding to
$\vec{p}/{m} \to \vec{p}/m + \delta \vec{j}_0$.
The energy correction reads as
\begin{align}
& \delta E_E = 
- \frac{2 \alpha}{3 \pi m^2} \,
\langle h_{M,0} \rangle  
\\[2ex]
& \; \times \int_0^{m \epsilon} d \omega \, \omega \,
\left< j \tfrac12 \left| \vec{p} \,
\left( \frac{1}{H_{\rm NR}  - E_{\rm NR}  + \omega} \right)^2 \,
\vec{p} \, \right| j \tfrac12 \right> \,,
\nonumber
\end{align}
where $m \epsilon$ is an upper cutoff for the photon 
energy~\cite{Be1947} (the scale-separation parameter 
$\epsilon$ is dimensionless).
It corresponds to the following $g$ factor correction,
\begin{equation}
\delta g_E(nP_j) = g_{00}(nP_j) \, 
\frac{\delta E_E}{\langle h_{M,0} \rangle} \,.
\end{equation}
A numerical evaluation of this correction according to 
established techniques~\cite{JePa1996} yields
\begin{subequations}
\label{gE}
\begin{align}
\delta g_E(2P_{1/2}) =& \; 
\frac{2 \alpha}{3 \pi} (Z\alpha)^2 
\left( -\frac16 \, \ln\left(\frac{\epsilon}{(Z\alpha)^2}\right) 
- 0.12831 \right) ,
\\[2ex]
\delta g_E(2P_{3/2}) =& \; 
\frac{4 \alpha}{3 \pi} (Z\alpha)^2 
\left( -\frac16 \, \ln\left(\frac{\epsilon}{(Z\alpha)^2}\right) 
- 0.12831 \right) .
\end{align}
\end{subequations}

For the correction to the Hamiltonian, we get
\begin{align}
& \delta E_H = 
\frac{2 \alpha}{3 \pi m^2} \,
\int_0^{m\epsilon} d \omega \, \omega \,
\nonumber
\\[2ex]
& \quad \times 
\left< j \tfrac12 \left| \vec{p} 
\left( \frac{1}{H_{\rm NR}  - E_{\rm NR} + \omega} \right)^2 \, 
h_{M,0} \, \vec{p} \, \right| j \tfrac12 \right> \,,
\end{align}
where we have used the relation $[H_{\rm NR}, h_{M,0}] = 0$.
This translates into the following correction for the 
$g$ factor,
\begin{equation}
\delta g_H(nP_j) = g_{00}(nP_j) \, 
\frac{\delta E_H}{\langle h_{M,0} \rangle} \,.
\end{equation}
A numerical evaluation leads to the following results,
\begin{subequations}
\begin{align}
\delta g_H(2P_{1/2}) =& \; \frac{2 \alpha}{3 \pi} (Z\alpha)^2 
\left( \frac16 \,
 \ln\left(\frac{\epsilon}{(Z\alpha)^2}\right) 
+ 0.25134 \right) \,,
\\[2ex]
\delta g_H(2P_{3/2}) =& \; \frac{4 \alpha}{3 \pi} (Z\alpha)^2 
\left( \frac16 \,
\ln\left(\frac{\epsilon}{(Z\alpha)^2}\right) 
+ 0.15907 \right) \,.
\end{align}
\end{subequations}

The correction to the current is given by
\begin{align}
\delta E_C =& \; - \frac{4 \alpha}{3 \pi m^2} \,
\int_0^{m\epsilon} d \omega \, \omega \,
\nonumber\\[2ex]
& \; \times \left< j \tfrac12 \left| p^i 
\frac{1}{H_{\rm NR}  - E_{\rm NR} + \omega}
\delta j^i_0
\right| j \tfrac12 \right> \,,
\end{align}
where we take into account the multiplicity factor
due to the current acting on both sides of the propagator.
The corresponding correction to the $g$ factor is
\begin{equation}
\delta g_C(nP_j) = g_{00}(nP_j) \, 
\frac{\delta E_j}{\langle h_{M,0} \rangle} \,.
\end{equation}
We obtain the following numerical results,
\begin{subequations}
\begin{align}
\delta g_C(2P_{1/2}) =& \; \frac{2 \alpha}{3 \pi} (Z\alpha)^2  \,\,
0.24607 \,,
\\[2ex]
\delta g_C(2P_{3/2}) =& \; \frac{4 \alpha}{3 \pi} (Z\alpha)^2  \,\,
0.06151 \,.
\end{align}
\end{subequations}
Summing all low-energy corrections, the spurious logarithmic
terms cancel, and we obtain
\begin{subequations}
\label{bL20}
\begin{align}
b^{(L)}_{20}(2P_{1/2}) =& \; 0.98428\,,
\\[2ex]
b^{(L)}_{20}(2P_{3/2}) =& \; 0.49216\,,
\end{align}
\end{subequations}
as the spin-dependent low-energy contribution to the bound-electron $g$ factor.
The result for $2P_{3/2}$ is equal to half the correction 
for $2P_{1/2}$ (this fact is independently proven also Appendix~\ref{appa}). 
We denote the low-energy contribution to the 
$b_{20}$ coefficient defined in Eq.~\eqref{defcoeff} by $b^{(L)}_{20}$.

%
%
\subsection{Length Gauge}
\label{lep2}

In the length gauge, the interaction with the 
quantized electromagnetic field is given by the 
dipole interaction $-e \, \vec{x}\cdot \vec{E}$,
where $\vec{E}$ is the electric-field operator~\cite{JeKe2004aop}.
The gauge-invariant~\cite{JeKe2004aop} nonrelativistic self-energy in the 
length-gauge reads
\begin{align}
\label{SElength}
\delta E =& \; 
- \frac{2 \alpha}{3 \pi} \,
\int_0^{m \epsilon} d \omega \, \omega^3 \,
\left< j \tfrac12 \left| \vec{x} \,
\frac{1}{H_{\rm NR}  - E_{\rm NR}  + \omega} \,
\vec{x} \right| j \tfrac12 \right> .
\end{align}
In the length gauge, the contribution of the 
wave-function correction vanishes because of the same reasons
as for the velocity gauge. 
Also, there is no correction to the transition current,
because the canonical momentum does not enter the interaction Hamiltonian 
in the length gauge.
We only have corrections to the Hamiltonian and to the energy. 
We start with the energy perturbation,
\begin{align}
& \delta E_{\mathcal E} = - \frac{2 \alpha}{3 \pi} \,
\langle h_{M,0} \rangle 
\\[2ex]
& \; \times \int_0^{m\epsilon} d \omega \, \omega^3 \,
\left< j \tfrac12 \left| \vec{x} \,
\left( \frac{1}{H_{\rm NR}  - E_{\rm NR}  + \omega} \right)^2 \,
\vec{x} \right| j \tfrac12 \right> \,.
\nonumber
\end{align}
The subscript $\mathcal E$ instead of $E$ serves to differentiate
the length-gauge as opposed to the velocity-gauge form of the correction.
Indeed, the 
numerical results for the 
corresponding correction to the 
bound-electron $g$ factor are different from those given in Eq.~\eqref{gE} and read
\begin{subequations}
\begin{align}
\delta g_{\mathcal E}(2P_{1/2}) =& \; \frac{2 \alpha}{3 \pi} (Z\alpha)^2 
\left( -\frac12 \, \ln\left(\frac{\epsilon}{(Z\alpha)^2}\right) 
- 0.88488 \right) \,,
\\[2ex]
\delta g_{\mathcal E}(2P_{3/2}) =& \; \frac{4 \alpha}{3 \pi} (Z\alpha)^2 
\left( -\frac12 \,
\ln\left(\frac{\epsilon}{(Z\alpha)^2}\right) 
- 0.88488 \right) \,.
\end{align}
\end{subequations}

For the correction to the Hamiltonian, we get
\begin{align}
& \delta E_{\mathcal H} =
\frac{2 \alpha}{3 \pi} \,
\int_0^{m \epsilon} d \omega \, \omega^3 \,
\nonumber\\[2ex]
& \; \times \left< j \tfrac12 \left| \vec{x} 
\left( \frac{1}{H_{\rm NR}  - E_{\rm NR} + \omega} \right)^2 \, h_{M,0} \,
\vec{x} \, \right| j \tfrac12 \right> \,.
\end{align}
A numerical evaluations leads to
\begin{subequations}
\begin{align}
\delta g_{\mathcal H}(2P_{1/2}) =& \; \frac{2 \alpha}{3 \pi} (Z\alpha)^2 
\left( \frac12 \, \ln\left(\frac{\epsilon}{(Z\alpha)^2}\right) 
+ 1.25399 \right) \,,
\\[2ex]
\delta g_{\mathcal H}(2P_{3/2}) =& \; \frac{4 \alpha}{3 \pi} (Z\alpha)^2 
\left( \frac12 \,
\ln\left(\frac{\epsilon}{(Z\alpha)^2}\right) 
+ 0.97716 \right) \,.
\end{align}
\end{subequations}
Adding the length-gauge corrections, the logarithmic terms cancel,
and it is straightforward to numerically
verify the gauge-invariance relation
\begin{equation}
\delta g_E + \delta g_H + \delta g_C =
\delta g_{\mathcal E} + \delta g_{\mathcal H} 
\end{equation}
and thus, the numerical results already given in Eq.~\eqref{bL20}.

Let us finally discuss the analytic proof of the gauge invariance.
Using the commutator relation
\begin{equation}
\frac{p^i}{m} = \ii \, [H_{\rm NR} - E_{\rm NR} + \omega, \, x^i], 
\end{equation}
and with the help of a somewhat lengthy calculation,
it is possible to show analytically that the 
velocity-gauge and the length-gauge forms of the 
low-energy contributions are equal. The calculation
follows ideas outlined in detail in Ref.~\cite{WuJe2009} 
where the more complicated case of a relativistic 
correction to a transition matrix element was considered.
Here, we are interested mainly in the numerical value of the
correction, for which the gauge invariance provides a
highly nontrivial check. Note that the matrix
elements governing the transitions of the reference to the 
virtual states are completely different in the length and 
in the velocity gauges, and the final results
are obtained after summing over the discrete and continuous 
parts of the spectrum of virtual states.

%
%
\section{Conclusions}
\label{conclu}

In our approach to the calculation of the 
bound-electron $g$ factor of $P$ states,
the contribution due to high-energy virtual photons can be
obtained using two alternative approaches,
based either on an effective Dirac equation or
on a low-energy effective Hamiltonian.
The contribution due to low-energy photons is treated as a
perturbation of the bound-electron self-energy~\cite{Be1947}
due to the interaction with the external uniform 
magnetic field.
Corrections to the Hamiltonian, to the bound-state energy
and (in the velocity gauge) to the transition current 
have to be considered.
The final results for the low-energy parts in the 
velocity- and length-gauges agree
although the individual contributions differ (including the
coefficients of spurious logarithmic terms).

Adding the high-energy contribution to the 
$g$ factor correction given in Eq.~\eqref{bH20}
and the low-energy effect given in Eq.~\eqref{bL20},
we obtain the following results for the 
self-energy correction of order $\alpha (Z\alpha)^2$ 
to the bound-electron $g$ factor of $2P$ states
($b_{20} = b^{(H)}_{20} + b^{(L)}_{20}$),
\begin{subequations}
\label{final}
\begin{align}
b_{20}(2P_{1/2}) =& \; 0.48429\,,
\\[2ex]
b_{20}(2P_{3/2}) =& \; 0.59214\,,
\end{align}
where the $b_{20}$ coefficient has been defined in Eq.~\eqref{defcoeff}.
Both above results compare favorably with 
recently obtained numerical data for low-$Z$ hydrogenlike
ions~\cite{YeJe2010} (see also Appendix~\ref{appb}).
An obvious generalization of the formalism outlined here
to the $3P$ and $4P$ states yields the results
\begin{align}
b_{20}(3P_{1/2}) =& \; 0.40500\,,
\\[2ex]
b_{20}(3P_{3/2}) =& \; 0.55250\,,
\\[2ex]
b_{20}(4P_{1/2}) =& \; 0.31331\,,
\\[2ex]
b_{20}(4P_{3/2}) =& \; 0.50665\,.
\end{align}
\end{subequations}

The two main results of the current investigation 
can be summarized as follows.
First, in Sec.~\ref{lep} we formulate a generalizable procedure for the 
calculation of low-energy corrections to the Land\'{e} $g_j$
factors in one-electron ions, applicable to $P$
states and states with higher angular momenta.
This procedure is based on choosing a specific 
reference axis for the external magnetic field.
In the future, it might be applied to include higher-order terms from 
the Hamiltonian~\eqref{Hmag} which couple the 
orbital and spin degrees of freedom.
Second, we resolve the discrepancy reported in Ref.~\cite{YeJe2010}
regarding the low--$Z$ limit of the $\alpha \, (Z\alpha)^2$ 
correction to the $g_j$ factor with previous 
results reported in Ref.~\cite{CaGrOw1978} for this correction
[see Eq.~\eqref{final} and Appendix~\ref{appb}].
In Appendix~\ref{appa}, it is shown that the discrepancy to 
the results of Ref.~\cite{CaGrOw1978} can be traced to 
the final evaluation of the logarithmic sums over virtual states,
while the angular momentum algebra is in agreement.
That is a further reason why 
the cross-check of our calculation in the length and velocity gauges
appeared to be useful.

Regarding the experimental usefulness of the obtained 
results, we can say that recent proposals~\cite{QuNiJe2008}
concerning measurements of the bound-electron $g$ 
factor of low--$Z$ hydrogenlike ions 
are based on double-resonance schemes that also involve
transitions to $2P$ states in the presence of the strong magnetic 
fields of Penning traps. In order to fine-tune the double-resonance
setup, the results obtained here might be useful.
Also, the results reported here serve as a general 
verification for the analytic formalism used in the 
theoretical treatment of $\alpha^3$ corrections to the $g$ factor of $2^3 P$ states in helium,
for which an interesting discrepancy of experimental 
and theoretical results persists (see Ref.~\cite{LhFaBi1976}
and Sec.~V of Ref.~\cite{Pa2005}).

We conclude with the following remark.  Our calculation concerns the $g$ factor
of $P$ states, and we confirm that in the order $\alpha (Z\alpha)^2$, low-energy
virtual photons yield an important contribution.  From the treatment in
Sec.~\ref{lep}, we can understand physically why there is no such low-energy effect
of order $\alpha(Z\alpha)^2$ for $S$ states.  Namely, a Schr\"{o}dinger--Pauli
$S_{1/2}$ state happens to be an eigenstate of the Hamiltonian $h_{M,0}$.
We have $h_{M,0} | nS_{1/2} (\mu = \pm \tfrac12 ) \rangle = \pm  | nS_{1/2} (\mu =
\pm \tfrac12 ) \rangle$.  This property holds because an $S$
state carries no orbital angular momentum, and therefore is an eigenstate of
the third component of the spin operator $\sigma_0$,
and $b_{20}$ therefore vanishes for $S$ states~\cite{PaJeYe2004}. 
For $P$ states and states with higher orbital angular momenta, 
the situation is different:
these states are not eigenstate of $h_{M,0}$ irrespective of 
their angular momentum projection, even though $h_{M,0}$ commutes with the
nonrelativistic Hamiltonian $H_{\rm NR}$.  Therefore, there is a residual
effect of order $\alpha (Z\alpha)^2$ due to low-energy virtual photons
for states with nonvanishing angular momenta.

%
%
\section*{Acknowledgments}

Valuable and insightful discussions with V.~A.~Yerokhin and K.~Pachucki are
gratefully acknowledged.  This project was supported by the National Science
Foundation (Grant PHY--8555454) and by a precision measurement grant from the
National Institutes of Standards and Technology (NIST).

\appendix

\begin{table}[htb]
\caption{\label{table1} Higher-order remainder functions
$i_{1/2}(Z\alpha)$ and $i_{3/2}(Z\alpha)$ for the 
self-energy correction to the $g$ factor of $2P_{1/2}$
and $2P_{3/2}$ states, respectively, as obtained
recently in Ref.~\cite{YeJe2010}. The value of $\alpha$
employed in the calculation is $\alpha^{-1} = 137.036$, and
the numerical uncertainty of the all-order (in $Z\alpha$)
calculation due to the finite number of integration
points in the numerical calculation is indicated in brackets.}
\begin{tabular}{c@{\hspace{0.4cm}}.@{\hspace{0.4cm}}.}
\hline
\hline
$Z$ &
\multicolumn{1}{c}{$g_{1/2}(Z\alpha)$} &
\multicolumn{1}{c}{$g_{3/2}(Z\alpha)$} \\
\hline
1 &  0.121x258\,(21) &  0.148x104\,(21) \\
2 &  0.121x715\,(22) &  0.148x294\,(24) \\
3 &  0.122x414\,(19) &  0.148x567\,(24) \\
4 &  0.123x280\,(14) &  0.148x851\,(22) \\
5 &  0.124x305\,(10) &  0.149x338\,(18) \\
6 &  0.125x473\,(7)  &  0.149x816\,(14) \\
7 &  0.126x803\,(5)  &  0.150x350\,(11) \\
8 &  0.128x186\,(4)  &  0.150x933\,(7) \\
9 &  0.129x711\,(2)  &  0.151x561\,(4) \\
10&  0.131x336\,(2)  &  0.152x231\,(4) \\
\hline
\hline
\end{tabular}
\end{table}

\begin{figure}[thb]
\includegraphics[width=0.7\linewidth]{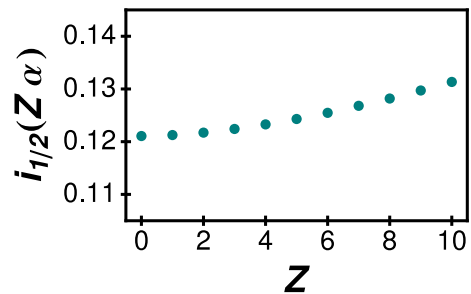}
\caption{\label{fig1} (Color online) The higher-order
remainder function $i_{1/2}(Z\alpha)$ is shown as a function
of $Z$. Numerical values for $i_{1/2}(Z\alpha)$
are given in Table~\ref{table1}. The point at $Z=0$
is given by the coefficient $\tfrac14 b_{20}(2P_{1/2})$
and is approached smoothly.}
\end{figure}

\begin{figure}[thb]
\includegraphics[width=0.7\linewidth]{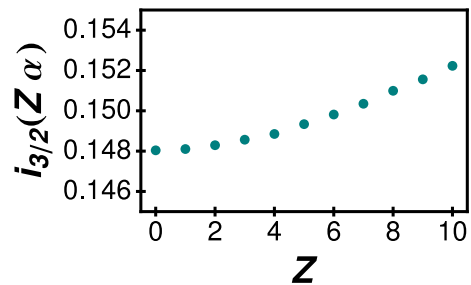}
\caption{\label{fig2} (Color online) Same as Fig.~\ref{fig1}, but for the
$2P_{3/2}$ state.  The higher-order
remainder function $i_{3/2}(Z\alpha)$ is plotted as a function
of $Z$, with numerical values for $i_{3/2}(Z\alpha)$
given in Table~\ref{table1}. The point at $Z=0$
is $\lim_{Z\alpha \to 0} i_{3/2}(Z\alpha) = \tfrac14 b_{20}(2P_{3/2})$.
The limit is approached smoothly.}
\end{figure}

%
%
\section{Remarks on the Low--Energy Part}
\label{appa}

First of all, let us remark that our results 
reported in Sec.~\ref{lep} can be expressed as 
logarithmic sums over the spectrum of atomic hydrogen.
After performing the angular algebra~\cite{VaMoKh1988},
we can write the total low-energy 
correction $\delta g^{(L)}(nP_j) = 
\delta g_{\mathcal E}(nP_j) +
\delta g_{\mathcal H}(nP_j)$ as follows,
\begin{align}
& \delta g^{(L)}(nP_{1/2}) = 
\frac{8 \alpha}{9 \pi} 
\sum_m (E_{mS} - E_{nP})^2 \,
\left< mS || \vec{r} || nP \right>^2 
\nonumber\\[2ex]
& \times \ln\left(\frac{2 | E_{mS} - E_{nP}|}{(Z\alpha)^2 m}\right) 
- \frac{8 \alpha}{9 \pi} \sum_m (E_{mD} - E_{nP})^2  \, 
\nonumber\\[2ex]
& 
\times \left< nP || \vec{r} || nD \right>^2 \,
\ln\left(\frac{2 | E_{mD} - E_{nP}|}{(Z\alpha)^2 m}\right) 
\end{align}
and 
\begin{align}
& \delta g^{(L)}(nP_{3/2}) = 
\frac{4 \alpha}{9 \pi} 
\sum_m (E_{mS} - E_{nP})^2 \,
\left< mS || \vec{r} || nP \right>^2 
\nonumber\\[2ex]
& \times \ln\left(\frac{2 | E_{mS} - E_{nP}|}{(Z\alpha)^2 m}\right) 
- \frac{4 \alpha}{9 \pi} \sum_m (E_{mD} - E_{nP})^2  \, 
\nonumber\\[2ex]
& 
\times \left< nP || \vec{r} || nD \right>^2 \,
\ln\left(\frac{2 | E_{mD} - E_{nP}|}{(Z\alpha)^2 m}\right) \,.
\end{align}
Here, $\left< mS || \vec{r} || nP \right>$ and
$\left< mD || \vec{r} || nP \right>$ are reduced matrix elements in the 
notation of Ref.~\cite{Ed1957}.
Because $(E_{mS} - E_{nP}) \propto (Z\alpha)^2$ and 
$\left< mS || \vec{r} || nP \right> \propto (Z\alpha)^{-1}$,
the above corrections to the $g$ factor are manifestly of order
$\alpha (Z\alpha)^2$. The sums over $m$ extend over both the 
discrete as well as the continuous part of the hydrogen spectrum
and can conveniently be evaluated using basis-set techniques~\cite{SaOe1989}.

Because it may not be completely evident from the presentation
in Ref.~\cite{CaGrOw1978}, we reemphasize here that
the authors of the cited article evaluate
a correction $\delta g_\ell$ to the orbital $g$ factor 
according to the definition
\begin{equation}
\label{gSgL}
H_{\rm eff} = 
g_s \, \mu_B \, \vec{S} \cdot \vec{B} +
g_\ell \, \mu_B \, \vec{L} \cdot \vec{B}
\end{equation}
for the effective interaction of a bound electron with an external
magnetic field ($\vec{S} = \tfrac12 \vec{\sigma}$ measures
the electron spin, and we have $g_S \approx 2$, $g_\ell \approx 1$).
Here, an interesting analogy to the description hyperfine splitting 
can be drawn, because the interaction with the nuclear magnetic 
field can also be separated into distinct components, namely 
the orbital, spin-dipole, and Fermi contact terms given, e.g.,
in Eqs.~(24)---(33) of Ref.~\cite{Ye2008}.

Returning to the discussion of the $g$ factor, 
we see that as a combination of $g_s$ and $g_\ell$, 
the Land\'{e} $g$ factor is obtained as

\begin{align}
\label{conv}
g_j =& \; g_\ell \,
\frac{j (j+1) + \ell (\ell + 1) - s(s+1)}{2 j(j+1)} 
\nonumber\\[2ex]
& \; + g_s \,
\frac{j (j+1) + s (s+1) - \ell (\ell + 1)}{2 j(j+1)} 
\nonumber\\[2ex]
\approx & \; \frac{3 j (j+1) - \ell (\ell + 1) + s(s+1)}{2 j(j+1)} \,.
\end{align}
In our notation, the analytic result given in Eq.~(14) of Ref.~\cite{CaGrOw1978}
for the correction to the orbital $g$ factor reads
\begin{align}
& \delta g_\ell(nP) = 
\frac{\alpha}{3 \pi} 
\sum_m (E_{mS} - E_{nP})^2 \,
\left< mS || \vec{r} || nP \right>^2 
\nonumber\\[2ex]
& \times \ln\left(\frac{2 | E_{mS} - E_{nP}|}{(Z\alpha)^2 m}\right) 
- \frac{\alpha}{3 \pi} \sum_m (E_{mD} - E_{nP})^2  \,  
\nonumber\\[2ex] 
& 
\times \left< nP || \vec{r} || nD \right>^2 \,
\ln\left(\frac{2 | E_{mD} - E_{nP}|}{(Z\alpha)^2 m}\right)  \,.
\end{align} 
The prefactors multiplying $g_\ell$ in 
Eq.~\eqref{conv} read $\tfrac43$ for $P_{1/2}$ 
and $\tfrac23$ for $P_{3/2}$ states, and therefore the 
analytic formula obtained in Eq.~(14) of Ref.~\cite{CaGrOw1978}
is in agreement with our approach for both $P_{1/2}$ and 
$P_{3/2}$ states, after the correction to $g_\ell$ is
converted into the corresponding modification of $g_j$.
However, their numerical result
$\delta g_\ell = -0.24 \alpha^3$ disagrees with our result
both in sign and in magnitude. 
Indeed, expressed in terms of our $b^{(L)}_{20}$ coefficient,
the results indicated in Ref.~\cite{CaGrOw1978} 
would imply that $b^{(L)}_{20}(2P_{1/2}) = -4.02$ and
$b^{(L)}_{20}(2P_{3/2}) = -8.04$.

Finally, let us note that the calculation of the 
orbital correction to the $g_j$ factor 
is only applicable to order $\alpha(Z\alpha)^2$, 
not $\alpha(Z\alpha)^4$, because
the higher-order terms in the 
magnetic interaction~\eqref{Hmag} couple 
the orbital and spin degrees of freedom.
The formalism outlined in Sec.~\ref{lep} generalizes
easily to the calculation of higher-order corrections 
that couple spin and orbital angular momentum,
which might be needed in the future,
whereas the separation into spin and orbital $g$ factors
only holds up to order $\alpha(Z\alpha)^2$.

%
%
\section{Comparison to Numerical Data}
\label{appb}

We parameterize the one-loop self-energy correction $\delta^{(1)} g$ to the 
$g$ factor of $P$ states as
\begin{align}
\delta^{(1)} g(2P_j) =& \; 
\frac{\alpha}{\pi} \left\{ b_{00} +  
(Z\alpha)^2 \,\, i_j(Z\alpha) \right\} \,,
\end{align}
where $i_j(Z\alpha)$ is the nonperturbative (in $Z\alpha$) 
remainder function. The remainder functions
$i_{1/2}(Z\alpha)$ and $i_{3/2}(Z\alpha)$ for $2P_{1/2}$ and 
$2P_{3/2}$, respectively, have recently been 
evaluated in Ref.~\cite{YeJe2010}.
Numerical values of $i_j(Z\alpha)$ for $Z = 1,\dots, 10$ 
are given in Table~\ref{table1}.
Note that these data imply a spin-dependence of the 
higher-order correction term $b_{20}$ beyond the 
spin-dependence of the high-energy part given by Eq.~\eqref{bH20}.
The limit as $Z\alpha \to 0$ of the remainder 
function $i_j(Z\alpha)$ is 
\begin{equation}
\lim_{Z\alpha \to 0} i_j(Z\alpha) =
\tfrac14 b_{20}(2P_j)\,.
\end{equation}
As shown in Figs.~\ref{fig1} and~\ref{fig2}, this limit is being consistently 
approached by the numerical data.


\begin{thebibliography}{10}

\bibitem{BrPa1967}
S.~J. Brodsky and R.~G. Parsons, Phys. Rev. {\bf 163},  134  (1967).

\bibitem{He1973}
R.~A. Hegstrom, Phys. Rev. A {\bf 7},  451  (1973).

\bibitem{GrKa1973}
H. Grotch and R. Kashuba, Phys. Rev. A {\bf 7},  78  (1973).

\bibitem{GrHe1973}
H. Grotch and R.~A. Hegstrom, Phys. Rev. A {\bf 8},  2771  (1973).

\bibitem{CaGrOw1978}
J. Calmet, H. Grotch, and D.~A. Owen, Phys. Rev. A {\bf 17},  1218  (1978).

\bibitem{AnSe1993}
J.~M. Anthony and K.~J. Sebastian, Phys. Rev. A {\bf 48},  3792  (1993).

\bibitem{GoSa1997}
I. Gonzalo and E. Santos, Phys. Rev. A {\bf 56},  3576  (1997).

\bibitem{YaDr1994}
Z.~C. Yan and G.~W.~F. Drake, Phys. Rev. A {\bf 50},  R1980  (1994).

\bibitem{Ya2002}
Z.~C. Yan, Phys. Rev. A {\bf 66},  022502  (2002).

\bibitem{Pa2004}
K. Pachucki, Phys. Rev. A {\bf 69},  052502  (2004).

\bibitem{LeHu1973}
M.~L. Lewis and V.~W. Hughes, Phys. Rev. A {\bf 8},  2845  (1973).

\bibitem{LhFaBi1976}
C. Lhuillier, J.~P. Faroux, and N. Billy, J. Phys. (Paris) {\bf 37},  335
  (1976).

\bibitem{JeEv2005}
U.~D. Jentschura and J. Evers, Can. J. Phys. {\bf 83},  375  (2005).

\bibitem{Br1928}
G. Breit, Nature (London) {\bf 122},  649  (1928).

\bibitem{ItZu1980}
C. Itzykson and J.~B. Zuber, {\em \relax{Quantum Field Theory}} (McGraw-Hill,
  New York, 1980).

\bibitem{Pa2005}
K. Pachucki, Phys. Rev. A {\bf 71},  012503  (2005).

\bibitem{VaMoKh1988}
D.~A. Varshalovich, A.~N. Moskalev, and V.~K. Khersonskii, {\em Quantum Theory
  of Angular Momentum} (World Scientific, Singapore, 1988).

\bibitem{Be1947}
H.~A. Bethe, Phys. Rev. {\bf 72},  339  (1947).

\bibitem{BiPaFFJo1995}
J. Biero\'{n}, F.~A. Parpia, C. Froese~Fischer, and P. J\"{o}nsson, Phys. Rev.
  A {\bf 51},  4603  (1995).

\bibitem{YeArShPl2005}
V.~A. Yerokhin, A.~N. Artemyev, V.~M. Shabaev, and G. Plunien, Phys. Rev. A
  {\bf 72},  052510  (2005).

\bibitem{JePa1996}
U. Jentschura and K. Pachucki, Phys. Rev. A {\bf 54},  1853  (1996).

\bibitem{JeKe2004aop}
U.~D. Jentschura and C.~H. Keitel, Ann. Phys. (N.Y.) {\bf 310},  1  (2004).

\bibitem{WuJe2009}
B.~J. Wundt and U.~D. Jentschura, Phys. Rev. A {\bf 80},  022505  (2009).

\bibitem{YeJe2010}
V.~A. Yerokhin and U.~D. Jentschura, Phys. Rev. A {\bf 81},  012502  (2010).

\bibitem{QuNiJe2008}
W. Quint, B. Nikoobakht, and U.~D. Jentschura, Pis'ma v ZhETF {\bf 87},  36
  (2008), [JETP Lett. {\bf 87}, 30 (2008)].

\bibitem{PaJeYe2004}
K. Pachucki, U.~D. Jentschura, and V.~A. Yerokhin, Phys. Rev. Lett. {\bf 93},
  150401  (2004), [Erratum Phys. Rev. Lett. {\bf 94}, 229902 (2005)].

\bibitem{Ed1957}
A.~R. Edmonds, {\em Angular Momentum in Quantum Mechanics} (Princeton
  University Press, Princeton, New Jersey, 1957).

\bibitem{SaOe1989}
S. Salomonson and P. \"{O}ster, Phys. Rev. A {\bf 40},  5559  (1989).

\bibitem{Ye2008}
V.~A. Yerokhin, Phys. Rev. A {\bf 78},  012513  (2008).

\end{thebibliography}
\end{document}